# Metrics to evaluate research performance in academic institutions: A critique of ERA 2010 as applied in forestry and the indirect $H_2$ index as a possible alternative

*Jerome K Vanclay*[1] and *Lutz Bornmann*[2]

1. *Southern Cross University, PO Box 157, Lismore NSW 2480, Australia. JVanclay@scu.edu.au*

2. *Max Planck Society, Administrative Headquarters, PO Box 101062, 80084 Munich, Germany*

**Abstract**

Excellence for Research in Australia (ERA) is an attempt by the Australian Research Council to rate Australian universities on a 5-point scale within 180 Fields of Research using metrics and peer evaluation by an evaluation committee. Some of the bibliometric data contributing to this ranking suffer statistical issues associated with skewed distributions. Other data are standardised year-by-year, placing undue emphasis on the most recent publications which may not yet have reliable citation patterns. The bibliometric data offered to the evaluation committees is extensive, but lacks effective syntheses such as the h-index and its variants. The indirect $H_2$ index is objective, can be computed automatically and efficiently, is resistant to manipulation, and a good indicator of impact to assist the ERA evaluation committees and to similar evaluations internationally.

**Introduction**

The Australian Government's ERA (Excellence in Research for Australia) is an attempt by the Australian Research Council (ARC) to rate the research impact of each university on a 5-point scale within 180 Fields of Research (ARC 2010a) using a combination of metrics and peer evaluation by an evaluation committee (2011c). The ERA has been controversial for several reasons, including its reliance on journal ranking (Cooper and Poletti 2011; Haddow and Genoni 2010; Moosa 2011; Northcott and Linacre 2010; Pontille and Torny 2010; Serenko and Dohan 2011), its use of grant income as an indicator (Svantesson and White 2009), its reliance on peer review in some disciplines (Abramo et al. 2009; Abramo and D'Angelo 2011; Taylor 2011), and the high cost (involving a budget allocation of AU$35.8 million; Carr 2009). In addition, there remain questions surrounding the utility of esteem indicators (Donovan and Butler 2007).

Journal rankings formed an integral part of ERA in 2010, but were omitted from the 2012 ERA (Atkinson and McLoughlin 2011; Runeson 2011) and replaced with a 'journal indicator list'. While this list overcomes some concerns associated with the journal ranking (Vanclay 2011, 2012), it introduces new issues: it is not clear how this list will be used by the Research Evaluation Committees

(RECs) who may be unfamiliar with the journals on the list, and overlooks the fact that all journals have good, bad and indifferent contributions (Singh et al. 2007). ARC 2012 guidelines state that "*The table will inform expert judgements regarding the relevance of the journals to the research being published e.g. 'Is this an appropriate journal for this research?'. 'Is it a highly regarded journal?' … REC members will be able to drill down to article level data from the table, so will not be making their judgments solely on the basis of journal titles or article counts.*" (ARC 2011a). However, because of the magnitude of the task, it is unclear whether RECs have the time or the expertise to deal adequately with the large number of papers presented for evaluation (over 333,000 research outputs were evaluated in ERA 2010; ARC 2011b). All these research outputs have already been read by reviewers, editors and citing authors – suggesting that citation analysis may be an efficient alternative to peer review (Bornmann 2011). Thus this paper explores the utility of a metric derived from Hirsch's (2005) h-index that may provide an efficient alternative (or complement) to the journal indicator list consistent with the goals of the ERA.

**Literature**

During its 30-year evolution, there has been a tendency for evaluation systems of university research to become more complex and expensive (Hicks 2009), and there is renewed interest in the use of metrics to streamline these evaluations (Moed 2009). In particular, Hirsch's (2005) h-index spawned great interest and a large number of variants and comparative studies (Alonso et al. 2009; Bornmann and Mutz 2008; Bornmann et al. 2011; Norris and Oppenheim 2010; Schreiber 2010). While each metric has attractions and limitations, one variant, the single-publication h-index (Schubert 2009; Egghe 2011; Bornmann et al. in press), is of particular interest as it focuses on the impact of a single paper, without regard for the apparent impact of the journal. An article has a single-publication h-index of $n$ if it is cited $n$ times by other articles, each of which is itself cited at least $n$ times. This indicator is preferable to simple citation counts, as it relies on substantive citations and is less prone to manipulation. The number of citations to a paper can be influenced easily, e.g., by self-citations, but the ability to influence citations of the citing papers (i.e., the single-publication h-index) is very limited. It is also close to the ideals of the ERA, as it seeks to evaluate a single research output.

The single-publication h-index is a specific variant of the successive h-index (Schubert 2007), which has received considerable attention, especially regarding the 'downstream' aspects of institutional (Arencibia-Jorge et al. 2008; Rousseau et al. 2010; Ruane and Tol 2008) and national standing (Egghe 2008). Thus a publication may accumulate $h_0$ citations; an author may attain a successive h-index of $h_1$ if they have sufficient well-cited publications, an institution of $h_2$ if they have sufficient well-cited scientists, a nation an index of $h_3$ if they have sufficient well-established institutions, and so on. At each stage, an index $h_{i+1}$ takes the value $n$ only if it comprises $n$ instances of $h_i \geq n$. Although much attention has been devoted to the downstream aspects, little attention has been devoted to the equivalent 'upstream' aspects. The upstream aspects, where the value attributed to a citation depends on how often that work has itself been cited, is implicit in the single-publication h-index, and is of interest in evaluating the impact of a body of work by an individual, research team or institution. Egghe (2011) has suggested that the 'upstream' or indirect h-index be denoted with capital H to discriminate it from other variants of the h-index. The indirect $H_1$-index is a citation count: an article has $H_1 = n$ if it is cited by $n$ other articles. An author or team has an indirect index $H_2$ of $n$ if they have published $n$ items, each of which has $H_1 \geq n$ (i.e., cited by $n$ other articles, each of which is itself cited at least $n$ times; Figure 1).

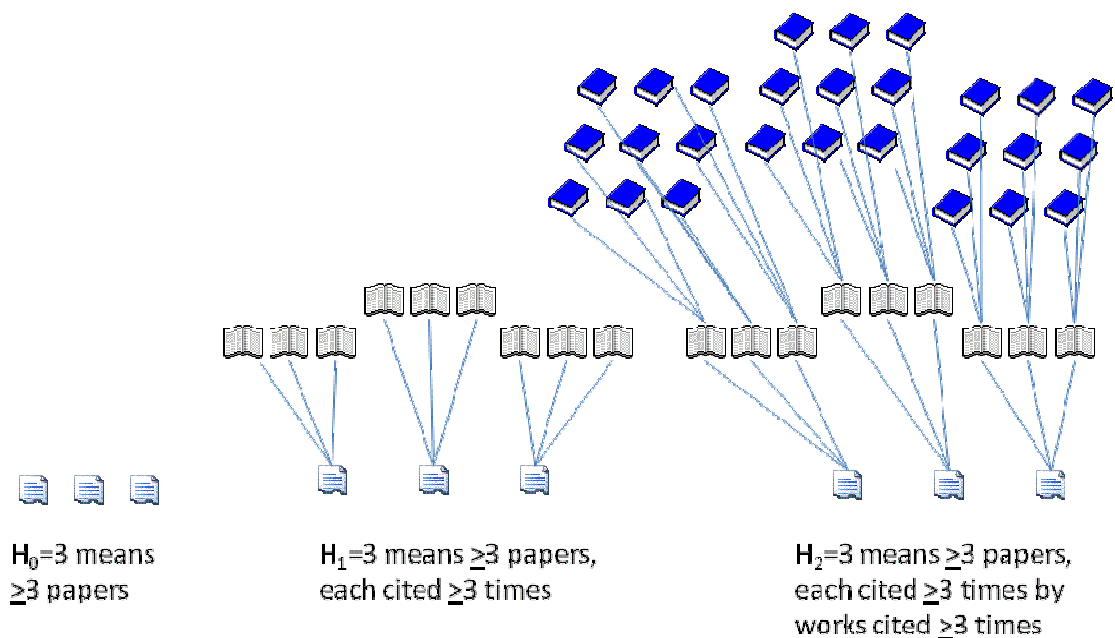

**Figure 1**. The derivation of $H_2$.

The use of the indirect index $H_2$ offers a stronger indication of citation impact: a publication count of *n* signifies output but not impact; a h-index of *n* (*n* publications, each cited *n* times) is a weak indicator of impact; a $H_2$ of *n* is a strong indicator impact because it assures that the citing articles are themselves cite-worthy. A high $H_2$ index of a publication typically indicates that papers associated with the research front in a field (the highly-cited papers) are based on this publication. Thus, the publication in question contributes to the scientific progress in the field (see Bornmann et al. 2010). The strength of $H_2$ is that it cannot be manipulated as easily as a citation count on which the conventional h-index relies (Labbé 2010). This is because $H_2$ relies on citations by substantive papers that are themselves heavily-cited, and overlooks less substantive citations irrespective of frequency. The use of $H_2$ rewards 'good' science: simple publication counts encourage publication without regard for impact; citation counts and the h-index reward provocative comment and literature reviews that may be often cited; while $H_2$ rewards foundation science that will be noticed in frequently-cited reviews. This is consistent with the stated aims of ERA to "[drive] research excellence rather than just research quantity" (Jones 2011). The $H_2$-index is easily computed using Scopus and Web of Science, and a web application is available to calculate the index with Google Scholar (Thor and Bornmann 2011). One limitation of $H_2$ is that it develops more slowly than more direct indicators such as citation counts. And like all such indicators, the quality of the $H_2$ estimates remains dependent on the quality of the underlying bibliometric database (Meho and Yang 2007).

The Australian ERA (Excellence in Research for Australia), like most research evaluations, seeks to appraise research achievements during a defined reference period – in contrast to the conventional application of the h-index which summarises life-time achievement. However, the h-index, the indirect $H_2$ and other related indices may be applied to defined periods such as the 6-year reference period used in the ERA (2003-2008 for ERA 2010, and 2005-2010 for ERA 2012).

The ERA seeks to evaluate research achievement in each of 180 defined Fields of Research (FoRs, ANZSRC 2008), defined largely by a prescribed list of journals (ARC 2010b). Derivatives of the h-index, including the indirect $H_2$ index, apply equally to such selected subsets as to the uncensored literature. This paper seeks to evaluate the utility of the $H_2$ index for automating an assessment of literature within FoRs, both generally and in the specific case of forestry (FoR 0705 Forestry

Sciences). The Australian Research Council's list of journals (ARC 2010b) that define this field has some limitations (Vanclay 2008, 2011), but it nonetheless provides a serviceable yardstick with which to compare institutions. Several aspects of this yardstick are examined in this paper.

The ERA rankings are allocated subjectively by Research Evaluation Committees guided by some bibliometric data, including tables showing the standing of institutions within each FoR based on relative citation impact (per article, 7 classes delimited by 0.01, 0.8, 1.2, 2, 4, 8 cites/paper; ARC 2011c) and on percentile analysis (5 classes, top 1%, 5%, 10%, 25%, 50% based on cites/paper; ARC 2011c). ERA 2010 also reported the mean relative citation impact (RCI) for each institution, so that an institution with 50 'good' publications is likely to be ranked higher than an institution with 100 'good' and 100 'average' publications. The disadvantage of RCI is that citation distributions are—as a rule—highly skewed and should not be arithmetically averaged. With percentile ranks, the citation of each paper is rated in terms of its percentile in the citation distribution (Leydesdorff et al. 2011). ERA's use of percentiles is commendable and has many advantages over other current standard techniques (based on mean citation rates). Specific RCIs and percentiles have not been disclosed by ERA, but comparable data for 0705 Forestry Sciences during 2005-10 are reported in this paper.

**Data**

The present study involves diverse sources of data:

1. the ERA rankings from 2010 (ARC 2011e),
2. the ARC list of journals classified as 0705 Forestry Sciences (Lamp 2011; Annex 1),
3. a compilation of forestry institutions that are active in research and training, and
4. bibliographic records retrieved from Elsevier's Sciverse Scopus, the official data provider to ERA in 2010.

Scopus provided many tools to facilitate the analysis, and most information was retrieved through a search defined by an affiliation-identifier, a list of ISSNs of journals, and the census period (2003-08 or 2005-2010). However, periodic updates of Scopus alter some data, and it is not possible for a casual user to specify a reference date, so the same search conducted on two successive days may, in some instances, return a different number of citations. To minimize the effects of such updates, efforts were made to retrieve relevant Scopus data for each comparison within a short time period, so that entries in any table remain comparable. A further complication is that the casual user of Scopus cannot derive retrospective values of $H_2$, because these may be inflated by recent references. Scopus provides some capacity to derive retrospective h-indices, but offers no ability for the casual user to limit citation counts at the second tier required to estimate $H_2$ (Figure 1).

A list of 380 forestry institutions (Annex 2) was compiled, initially from Wikipedia (Anon 2011) and Laband and Zhang (2006), but with additions and deletions to focus on tertiary educational institutions active in forest research and training. This list focuses on universities and other tertiary institutions engaged in both education and research, and omits government research institutions whose primary focus is research rather than training. Institutions with no published output during 2005-2010 in 0705 Forestry journals visible to Scopus are omitted from this list. The list should be viewed as representative rather than authoritative, because deficiencies in the 0705 list of journals may cause the omission of some worthy institutions, especially amongst those that publish in non-English journals. The 0705 list (Annex 1) of 85 journals recognised by the 2010 ERA does not include all relevant journals (Vanclay 2011) and has poor coverage of journals in languages other than English. Despite these limitations, it remains the yardstick defined by the Australian Research Council for use in the ERA.

**Method**

The overall objective of the present analysis is to examine the utility of the indirect $H_2$ index to calculate automatically the research performance of institutions without the use of journal rankings. In doing so, it explored 4 underlying questions:

1. How effective is the ERA journal list in defining a body of research?
2. How does $H_2$ compare with ERA ratings? (in forestry and generally)
3. How do global forestry institutions rate on this scale, and what is 'world standard'?
4. How to adjust $H_2$ for the size of institution?

Most of these questions can be addressed with data derived from Scopus, using the advanced search facility. Many searches were restricted to match a FoR-specific journal list, and this can be specified easily and unambiguously by specifying journal ISSNs. Similarly, Scopus has a useful ability to specify an affiliation identifier to uniquely identify institutions. Scopus offers the ability to limit searches, for instance to exclude articles published in 2011, but not the ability to exclude citations received during 2011, so the casual user of Scopus is limited in their ability to backdate analyses. Thus many of the analyses reported focus on the 2012 ERA reference period (2005-10) because the period of citation accrual (8-80 months) more closely corresponds to that experienced during the ERA evaluation. At the time of writing, publications from 2003-08 have accrued citations for 32-104 months, substantially longer than the interval prevailing at the time of the ERA analysis.

**Results**

*How well do journals defined as 0705 Forestry cover the forestry literature?*

The analysis reported in this paper follows ERA in relying on three assumptions that define the data under analysis: the use of Scopus as a source of citation data, the use of the ERA journal list within each discipline, and the use of the ANZ classification of science disciplines. Thus the utility of the present analysis depends in part on the extent to which the ERA 0705 journals (Annex 1) adequately represent the forestry literature. Scopus was the official data provider to ERA 2010, and in some instances provides more comprehensive coverage than Web of Science, and more reliable coverage than Google Scholar, so represents a good data source (Falagas et al. 2008, Li et al. 2010). Table 1 shows a summary of the 1412 articles found with a Scopus search for documents published with an Australian author during 2005-10 with the 'Forestry' in the title, abstract or keywords. These articles in Table 1 appeared in 389 journals, including 35 of the 85 journals in Annex 1. These documents may be included in ERA explicitly if the journal is classified 0705 Forestry (e.g., *Forest Ecology and Management*, and 84 other journals in Annex 1), implicitly if the journal is classified 07 Agriculture (e.g., *Agricultural and Forest Entomology*) or MD Multidisciplinary (e.g., *Science* and *Nature*), or excluded if the journal is classified otherwise. In addition, some journals receive more than one classification (e.g., *Agricultural and Forest Meteorology* classified as 0401 Atmospheric Sciences and 0705 Forestry Sciences), and ERA allows submitting institutions to nominate which of the two FoR codes should be assigned to the article. Not all journals indexed by Scopus are recognised by ERA, and of the 1412 articles reported in Table 1, only 1304 are recognised by ERA. Of these, 605 would be detected by a search based on 0705 journals, but since some journals have more than one classification, the number of articles included in a 'hand-crafted' 0705 submission could vary from 550 (the minimal case with articles in journals classified only as 0705) to 802 (all articles both explicitly and implicitly classified as 0705 Forestry). Table 1 is indicative, because not all forestry articles include the keyword 'forestry' (e.g., articles about wood science may not mention forestry), and not all articles containing the word 'forestry' are primarily about forestry science (e.g., an article about sediments affecting coral might mention casually forestry and agriculture as possible sources of sediment). Nonetheless, Table 1 offers some reassurance that the ERA classification recognises a substantial proportion of the forestry material indexed by Scopus. Note that this view based on articles is more favourable than a comparable view based on journals, because the leading journal (*Forest Ecology and Management*) carries 16% of the articles, and 16% of the journals carry only one article. Whilst Table 1 is indicative for forestry, different patterns may arise for other disciplines and FoRs. Further flexibility was introduced in the 2012 ERA by the 'reassignment exception', which allows "articles which have significant content (66% or more) that could best be described by a particular four-digit FoR code" to be re-assigned to that Field of Research (ARC 2011a).

**Table 1**. Distribution within FoR codes of articles retrieved with a Scopus search for 'Forestry' with an Australian author published during 2005-10. Bold numbers indicate 550 articles explicitly, and 802 (197+605) articles implicitly coded as 0705 Forestry.

| FoR code | Excluded from 0705 | Included implicitly | Included explicitly | Subtotal | Fields included |
|---|---|---|---|---|---|
| 02, 03, 04 | 83 | | 1 | 84 | Physical, chemical, earth sciences |
| 05, 06 | 209 | 89 | 32 | 330 | Environmental, biological sciences |
| 07 | 49 | 22 | **550** | 621 | Agricultural sciences |
| 01, 08 | 8 | | | 8 | Mathematics, informatics & computing |
| 9, 12 | 108 | | 6 | 114 | Engineering, built environment |
| 11 | 3 | | | 3 | Medical and health sciences |
| 13, 14, 15, 16 | 31 | | 16 | 47 | Social, behavioural, economic sciences |
| 18, 19, 21, 22 | 11 | | | 11 | Humanities and creative arts |
| MD | | 86 | | 86 | Multidisciplinary |
| Subtotal | 502 | **197** | **605** | 1304 | |
| Unlisted | 108 | | | | |
| Total | | | | 1412 | |

Forestry fares rather well in such a test (Table 1), because much of forestry research in Australia appears in journals classified unambiguously as 0705 Forestry Sciences. This is not the situation for all disciplines, and microbiology offers a striking contrast to forestry. A search of Scopus for scientific articles containing the word 'microbiology', with an Australian affiliation published during 2005-10 yields a total of 3552 articles in 582 journals. Of these 582 journals, 48 are amongst the 162 journals on the ERA list of 0605 Microbiology journals. As with the forestry example, the distribution of articles is skewed, with one journal (*Journal of Virology*) publishing 7% of the articles and 8% of the journals bearing only one article. Table 2 illustrates that the FoR 0605 Microbiology is defined more ambiguously within ERA, with many of the relevant journals having multiple FoR classifications. The major ambiguity is with 1108 Medical Microbiology, with over one-third of the articles in the 'microbiology' search receiving multiple classifications including both these classifications (FoR 0605 Microbiology and 1108 Medical Microbiology). Table 2 reveals that an automated search of Scopus can provide a good approximation of an institution's ERA submission in 0705 Forestry Sciences, because it may retrieve 46% of the relevant material, while the ERA 2010 rules allowed the institution to vary this content between 42% and 62% of this material. In contrast, multiple classifications of journals (especially between 0605 Microbiology and 1108 Medical Microbiology) mean that while an automated Scopus search for 0605 Microbiology journals will still retrieve 47% of the material, an institution may arrange their submission to include as little as 3% or as much as 64% of the material. The flexibility will increase with the 'reassignment exception' in ERA 2012. The ambiguity reported in Table 2 applies equally to the ability of the ERA to compute relative citation impact (RCI) and to compile percentile analyses. This suggests that some reappraisal of the units of evaluation within ERA may be warranted.

**Table 2**. Comparison of the ambiguity in journal classifications for two Fields of Research (0705 Forestry Sciences and 0605 Microbiology) based on percentages of articles published by Australian University researchers during 2005-2010.

| Required in an ERA submission? | 0705 Forestry | | | 0605 Microbiology | | |
|---|---|---|---|---|---|---|
| | Yes | No | Subtotal | Yes | No | Subtotal |
| Yes (Journal has single 4-digit FoR code) | **42%** | | | **3%** | | |
| Maybe (Journal has multiple FoR codes)† | 4% | 15% | **62%** | 44% | 17% | **64%** |
| No (Journal not entitled to this FoR code) | | 38% | 38% | | 36% | 36% |
| Subtotal | **46%** | 54% | 100% | **47%** | 53% | 100% |
| Unlisted | | 8% | | | 4% | |

† when a journal bears more than one FoR code, the author's institution may select the FoR code to be applied for ERA purposes.

*What is the basis for the ERA ratings for 0705 Forestry?*

ERA has published the 2010 methodology (ARC 2011c), but not the data used to derive the rankings of Australian forestry institutions. It is not possible for the casual user of Scopus to reconstruct the data used in ERA 2010, but comparable data for the ERA 2012 reference period (2005-10) can be compiled. ERA standardises citation counts to calculate a relative citation impact (RCI) by calibrating to the global average citations per publication (cpp) within each FoR. Table 3 illustrates the national and global benchmarks used to standardise and classify 0705 Forestry submissions to ERA, compiled for the period 2005-10.

**Table 3**. Mean citations per article and percentile thresholds for 0705 Forestry during 2005-10

| Year of publication | Mean Citations per article | | Global citation thresholds for percentile analysis | | | | |
|---|---|---|---|---|---|---|---|
| | Australia | Global | 1% | 5% | 10% | 25% | Median |
| 2005 | 9.5 | 9.5 | 55 | 30 | 22 | 13 | 6 |
| 2006 | 8.7 | 7.6 | 47 | 24 | 18 | 10 | 5 |
| 2007 | 6.7 | 5.4 | 32 | 18 | 13 | 7 | 3 |
| 2008 | 4.4 | 4.0 | 23 | 13 | 10 | 6 | 3 |
| 2009 | 3.0 | 2.3 | 14 | 8 | 6 | 3 | 1 |
| 2010 | 1.2 | 0.9 | 7 | 4 | 3 | 1 | 0 |

One question that arises from Table 3 and the way ERA 2010 used percentile analysis is whether the citation pattern created by a global aggregation reflects that of individual institutions, or whether institutions vary in their citation patterns. For instance, one might suspect that dedicated research institutions might exhibit a different pattern of citations than an extension agency: the former might have a more diverse pattern (some highly-cited successes, a few speculative proposals infrequently cited) than the latter which might be expected receive consistent but modest citations. However, this speculation is not well supported by the evidence. Two major universities with elite forestry programs (Swedish University of Agricultural Sciences SLU, University of British Columbia; Annex 2) exhibit citation patterns similar to the pattern accruing to global pooled publications, suggesting that the pooled global data is adequate to calibrate Australian university output. Few extension agencies publish prolifically enough to establish a reliable trend, but the pooled data from several extension agencies exhibit a pattern similar to that of the Center for International Forestry Research (CIFOR) and the Australian CSIRO (Figure 2).

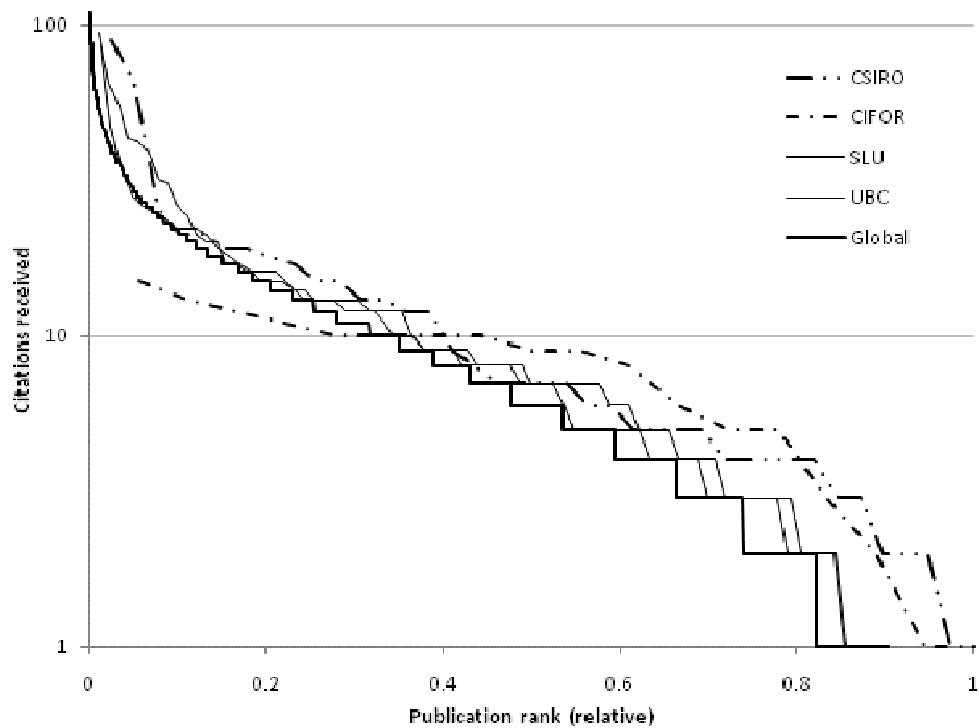

**Figure 2**. Citations to 0705 Forestry publications in 2005 pooled globally (thick solid line) compared with two leading forestry universities (UBC, University of British Columbia & SLU Swedish University of Agricultural Sciences, thin solid lines) and two research institutions (CSIRO, Commonwealth Scientific and Industrial Research Organisation & CIFOR Center for International Forest Research, dashed lines).

Although the evidence suggests that the pooled global data is a reasonable basis to standardize citation counts, it may not be desirable to do so. At the time of writing (August 2011), the mean citations per 2010 publication in 0705 Forestry was 0.9 (cpp, Table 3, global average), so a modest number of citations can easily inflate to high RCI (e.g., 8 citations is sufficient to attain the highest RCI class VI, Table 4). The threshold of the top percentile for 2010 publications is only 7 citations, so a modest cluster of citations (e.g., arising from a special issue of a journal) can very easily leverage an article into the top percentile. It seems imprudent to introduce such heavy weighting of the most recent publications, at a time when their long-term acceptance remains uncertain. Radicchi and Castellano (2011) show that for recent publication "citation patterns are generally far from stationary". Such weighting creates an unreasonable incentive for academics to indulge in citation manipulation, either by writing "potboilers" or by forming citation clubs.

A third problem with the data reported in Table 3 is that the mean citation counts (cpp) need to regarded cautiously, because the distribution of citations is highly skewed. The problem is evident internally in Table 3, since the mean values (left columns) differ greatly from the medians (right column). This difference is further illustrated in Figure 3 which reveals the skewed distributions, with few articles receiving over a hundred citations, and hundreds of articles receiving no citations. These distributions highlight the necessity for a percentile rank approach (rather than e.g. the mean RCI) in the analysis of research impact.

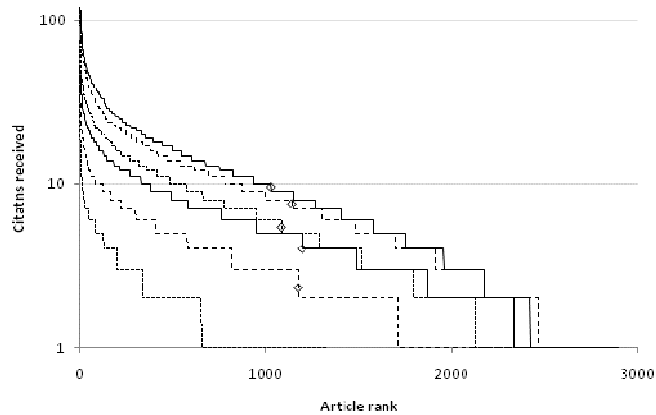

**Figure 3**. Citations accruing to 0705 Forestry articles globally during 2005-10. (By year: top line 2005; bottom line 2010). Diamond symbols indicate mean values.

The data presented in Table 3, plus the list of 0705 journals (Annex 1), allow the compilation of a table similar to that presented to Research Evaluation Committees (RECs) to assist them in their appraisal of institutions (ARC 2011c). Table 4 offers such a compilation for 0705 Forestry during 2005-10. The h-index and $H_2$ index are not included amongst the data presented to the RECs, but are included here as a possible synthesis or alternative. Our view is that Table 4 offers too much data and too little information, and that a more concise synthesis would be more insightful. It may be sufficient simply to present the percentiles in conjunction with the $H_2$.

**Table 4**. Example of data presented to Research Evaluation Committees for 0705 Forestry

| Performance indicator | | University of Tasmania | Southern Cross University | University of Melbourne | Australian National University | Australian average or total |
|---|---|---|---|---|---|---|
| Mean RCI | | 1.21 | 1.16 | 1.01 | 1.49 | 1.32 |
| RCI Class | RCI | Percent of outputs within RCI Class | | | | |
| 0 | 0 | 20% | 17% | 21% | 15% | 22% |
| I | 0.01-0.79 | 26% | 35% | 32% | 25% | 28% |
| II | 0.80-1.19 | 17% | 15% | 16% | 16% | 15% |
| III | 1.20-1.99 | 18% | 18% | 18% | 20% | 18% |
| IV | 2.00-3.99 | 13% | 11% | 10% | 17% | 12% |
| V | 4.00-7.99 | 6% | 5% | 4% | 8% | 6% |
| VI | ≥8.00 | 0% | 0% | 0% | 0% | 0.4% |
| Percentile | | Cumulative percent of outputs exceeding percentile | | | | |
| 1% | | 0% | 3% | 1% | 4% | 2% |
| 5% | | 9% | 8% | 5% | 10% | 8% |
| 10% | | 12% | 11% | 8% | 18% | 13% |
| 25% | | 43% | 35% | 35% | 48% | 39% |
| Median† | | 66% | 62% | 60% | 70% | 62% |
| Uncited articles | | 20% | 17% | 21% | 15% | 22% |
| Total indexed articles | | 116 | 66 | 135 | 89 | 835 |
| Hirsch h-index | | 10 | 10 | 11 | 15 | 28 |
| Indirect $H_2$ index | | 4 | 6 | 6 | 7 | 10 |
| ERA 2010 ranking | | 3 | 3 | 4 | 4 | |

† The median for 2010 articles includes 0 (Table3); these uncited articles have been omitted from this count.

Tables 1 and 2 have indicated some of the complications that arise as a result of the multiple classification of some journals. This problem has been exacerbated by the introduction of the 'reassignment exception', which allows great flexibility in the material that may be offered in an ERA submission. Table 5 indicates how, even in an unambiguously-specified field as forestry, this flexibility can lead to large changes in the data tabulated for the RECs, and reiterates the robustness and necessity of the h-index and $H_2$. Table 5 illustrates 3 cases: one that simply includes all material included in journals classified 0705 Forestry; another all-inclusive case that uses the reassignment exception to include all material with relevant keywords (forestry, silviculture, timber, eucalyptus), and a third selective case that uses the opportunity presented by the 'reassignment exception' to build the strongest possible case by omitting infrequently-cited material. It is noteworthy that the tabular REC data (RCI classes, percentiles) are easily manipulated in this way, but that the indirect $H_2$ index remains robust. The apparent three-fold increase in 'high impact outputs' (i.e., the proportion of work in RCI classes IV and V, and in top 10%) in the selective case is enabled through the flexibility of the 'reassignment exception', the ERA threshold of 50 articles, and use of means and percentiles.

**Table 5**. The 'reassignment exception' allows metrics to be manipulated – example with possible 0705 Forestry submissions from one institution

| Performance indicator | | All-inclusive | Strictly FoR 0705 | Highly selective |
|---|---|---|---|---|
| Mean RCI | | 1.23 | 1.16 | 2.26 |
| RCI Class | RCI | Percent of outputs within RCI Class | | |
| 0 | 0 | 17% | 17% | |
| I | 0.01-0.79 | 33% | 35% | |
| II | 0.80-1.19 | 15% | 15% | 24% |
| III | 1.20-1.99 | 15% | 18% | 32% |
| IV | 2.00-3.99 | 17% | 11% | 36% |
| V | 4.00-7.99 | 4% | 5% | 8% |
| Percentile | | Cumulative outputs exceeding percentile | | |
| 1% | | 3% | 3% | 6% |
| 5% | | 8% | 8% | 18% |
| 10% | | 16% | 11% | 34% |
| 25% | | 37% | 35% | 80% |
| Median | | 62% | 62% | 100% |
| Uncited articles | | 17% | 17% | |
| Total indexed articles | | 107 | 66 | 50 |
| Hirsch h-index | | 14 | 10 | 14 |
| Indirect $H_2$ index | | 6 | 6 | 6 |

*How does $H_2$ compare with ERA ratings?*

ERA attempts to calibrate the performance of universities in each of several Fields of Research (FoRs, ANZSRC 2008) defined via a prescribed set of journals (ARC 2010b). Most of the journal output for each institution can be retrieved from Scopus, because journal articles usually indicate an author's institutional affiliation. However, the journal classification is not unambiguous, with some journals assigned to more than one FoR (Tables 1 & 2), and this limits the ability to automate the computation of $H_2$ and other indicators of impact. In addition, the ERA ranking accounts for the publications of all staff within an institution at the census date (31 March of the preceding year), including newly-appointed staff whose earlier publications may not bear their new institution's name. Thus an automatically-compiled assessment from Scopus could differ from an institution's submission in both staff composition and in the selection of articles in multi-classified journals. Notwithstanding these differences, a Scopus compilation should offer a good approximation of ERA submissions by established institutions.

The basis for ERA rankings in 2010 varied between disciplines. In most areas of science considered in this paper, rankings were based on research outputs (ranked journals, citation analysis), grant income, esteem (membership of learned societies, research fellowships) and applied measures (patents, plant breeder's rights). Research outputs were assessed over a six year reference period (2003-08 in ERA 2010), whilst other indicators were assessed over a three year period (2006-08). It is likely that these indicators were correlated, since for instance, publications are a pre-requisite for grant success, but no data on these correlations and the implied redundancy have been made public. Thus a citation analysis drawn from Scopus (or other data providers) is unlikely to explain fully the ERA ratings, but it may offer some useful insights.

Indirect $H_2$ indices were computed for 0705 Forestry (including all journals classified 0705, but none classified 07 or MD) for the four institutions that received ERA ratings in 2010 (Table 4). Whilst the results show promise of a strong correlation between $H_2$ and ERA ranking, the sample size is too small for definitive conclusions.

Because the eligible forestry literature is unambiguously defined in the ERA rules (Table 1), it is feasible to automate a Scopus search to compute $H_2$ indices to assess performance of the institutions involved. However, forestry appears unique in this regard, and the ambiguous definition of the eligible literature in many other disciplines (Table 2), coupled with the other indicators involved (grant income, esteem, applied measures) means that automated estimates of $H_2$ do not correlate well with the ERA ranking of institutions (Table 6). However, Table 6 may offer an unduely pessimistic view, because it samples ill-defined FoRs (cf. Table 2), and compares ERA rankings from the reference period 2003-08 with $H_2$ indices derived for 2005-10 (due to limitations of Scopus). It appears desirable to re-examine both the scope of the Fields of Research and the FoRs assigned to journals appears needed, to limit 'game-playing' (e.g., Bornmann 2010) by institutions preparing submissions, and to allow the ERA evaluation process to be streamlined and partially automated.

**Table 6**. Confusion matrix contrasting classification by ERA 2010 rankings (2003-08) and by $H_2$ indices (2005-10) computed from Scopus on the basis of journal FoR classifications. Table entries reveal number of institutions within each class. Correct classifications are indicated in *italics*.

| ERA ranking | 1105 Dentistry | | | | 0912 Materials Science | | | 1115 Pharmacology | | | | 0608 Zoology | | |
|---|---|---|---|---|---|---|---|---|---|---|---|---|---|---|
| | 4 | 5 | 6-7 | 8 | ≤6 | 7-8 | ≥9 | 3-6 | 7 | 8-13 | ≥14 | 4-6 | 7-8 | >9 |
| 1 | | | | | | | | 2 | | | | | | |
| 2 | *1* | | | | *3* | | | *2* | 1 | | | 2 | 2 | |
| 3 | | *1* | | | 1 | *5* | | | *1* | 2 | | *3* | 2 | 1 |
| 4 | | | *2* | | 1 | | 5 | 2 | 1 | *4* | | 2 | *2* | |
| 5 | 1 | | | *1* | | 1 | 2 | | | 2 | *2* | 1 | 1 | *3* |
| Correct $H_2$ | 83% | | | | 72% | | | 47% | | | | 42% | | |
| Correct h-index | 50% | | | | 67% | | | 58% | | | | 53% | | |

It is noteworthy in Table 6 that different disciplines have different citation patterns, so that a $H_2$ value of 5 in 1105 Dentistry and 0608 Zoology corresponds to an ERA rating of 3 ("At world standard"), whereas a $H_2$ of 7 is required to achieve the same rating in 0912 Materials Science and 1115 Pharmacology. Such differences are commonly observed: for instance, Slyder et al. (2011) reported that forestry articles received more citations than geography articles from the same institution. Table 6 also reveals that $H_2$ performs better than other indicators (such as h-index) in well-defined FoRs (such as 1105 Dentistry with 83% correct) than in ill-defined FoRs (such as 0608 Zoology).

*How do forestry institutions compare on this scale?*

Forestry institutions worldwide attained $H_2$ scores between 0 and 11 during the 6-year period 2005-10 (Figure 4; Annex 2). The distribution of these scores is skewed, with many low scores, and few high scores. This distribution suffers censorship of three kinds: it includes only articles in 0705 Forestry journals (*sensu stricto*, Annex 1) and overlooks work in other journals (e.g., *Nature* and *Science*), some of which may be highly-cited, and so under-represents work of some strong institutions. It is biased towards Anglophone institutions, because of the English-language bias in Scopus (Moya-Anegón et al. 2007) and in the ERA 0705 list of journals (Annex 1). It focuses on institutions with greater research activity, and does not attempt a complete survey of all institutions, so under-represents institutions with low research output. Nonetheless, it offers an indicative distribution of institutional performance. Such distributions pose some challenges, in that that the mean (e.g., relative citation index) is meaningless, but value of percentiles (i.e., the 1%, 5%, 10%, 25% and 50% reference points used by ERA 2010) remain relatively robust, because of the integer nature and limited range of $H_2$.

Table 7. Percentiles of $H_2$ indices of forestry institutions.

| Percentile based on $H_2$ | $H_2$ | Selected institutions at or above percentile |
|---|---|---|
| 1% | 8 | Oregon State University<br>Swedish University of Agricultural Sciences<br>University of British Columbia |
| 5% | 7 | Australian National University<br>University of Washington<br>Wageningen University, Netherlands |
| 10% | 6 | Southern Cross University, Australia<br>University of Melbourne, Australia<br>University of Natural Resources and Life Sciences, Vienna<br>Virginia Tech, USA |
| 25% | 4 | Auburn University, USA<br>TU Dresden, Germany<br>University of Tasmania, Australia |
| Median | 3 | Bejing Forestry University, China<br>Cambridge University, UK<br>Czech University of Life Sciences, Prague |

The $H_2$ index corresponds well with other rankings of forest research institutions. Laband and Zhang (2006) ranked 53 North American university programs in forestry, basing their rankings on publications (both in total, and adjusted per full-time faculty), citations and reputation. Despite differences in methodology and the passage of time, the present $H_2$ index (Annex 2) exhibits a correlation of -0.73 (P<0.001) with the citation-based ranking of Laband and Zhang (2006).

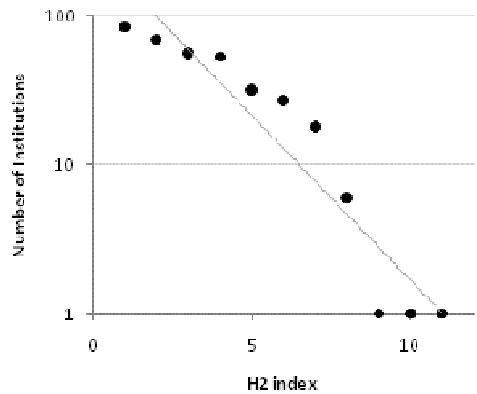

**Figure 4**. $H_2$ scores for universities worldwide

There is a high correlation between the $H_2$ index and the conventional h-index (0.95, P<0.001, Figure 5), so the question arises whether the established h-index (which is easier and more convenient to derive) has the same utility as the $H_2$ index. Similar results are reported by Bornmann et al (in press). However, Figure 5 illustrates that the same $H_2$ may have a wide range of h-index, and that the same h-index may give rise to a several different $H_2$ indices. For instance the following eight institutions have the same $H_2$ (6) but different h-index: Universität für Bodenkultur Wien (University of Natural Resources and Life Sciences, Vienna, h=17), Göttingen (16), Virginia Tech (15), Eidgenössische Technische Hochschule Zürich (Swiss Federal Institute of Technology, 14), Katholieke Universiteit Leuven (13), Freiburg (12), Cornell (11), Padova (10) – and these institutions are arguably of comparable standard with regard to their forestry research impact. Four other institutions have the same h-index (14), but different $H_2$: New Hampshire ($H_2$=8), Yale (7), Eidgenössische Technische Hochschule Zürich (6), Santiago de Compostella (Mexico, 5) – and these are arguably ranked in order of impact in forest research output. If it is accepted that the former eight are comparable, and the latter four are ranked, then the implication is that the indirect $H_2$ discriminates more effectively than the conventional h-index, at least in terms of the impact achieved by these institutions.

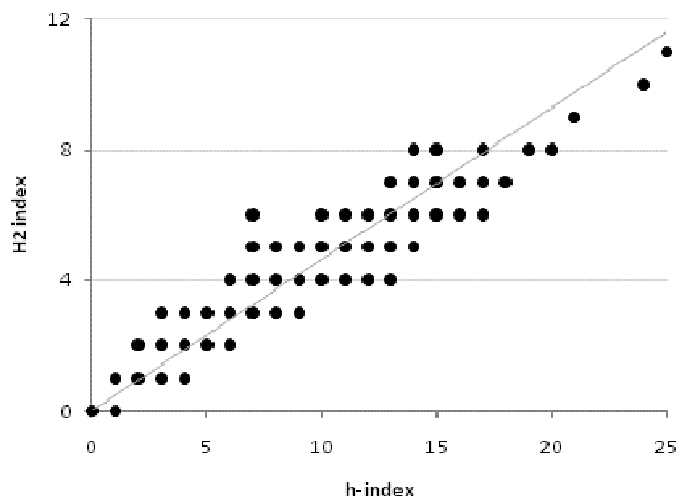

**Figure 5**. Comparison of h-index and $H_2$.

*How to adjust for size?*

Despite its stated aim to "[drive] research excellence rather than just research quantity" (Jones 2011), ERA appears to favour large institutions. In many evaluation systems including ERA and other university ranking systems, the top ranks are occupied by institutions that are both good and large, whereas the mid-ranks may be populated by a diverse mix of entities that are either good or large. Thus one of the challenges of an evaluation system is to discriminate between the often confounding effects of quality (citation impact) and quantity (publication output). The h-index and $H_2$ are both influenced to some extent by the number of contributors, so it is interesting to use the present data to explore the effect of size (Figure 6). Outputs for four forestry institutions have been reported above, based on numbers of research staff at these institutions and nationally reported by Sinclair (2010). The single-author datum in Figure 6 reflects the greatest $H_2$ of single-authored papers. These data suggest a logarithmic relationship between staff numbers and $H_2$ (Figure 6), with the line of best fit $H_2 = 3.0 + 1.05 \text{Ln}(N)$, suggesting that one way to adjust for research team size may be $H_2 - \text{Ln}(N)$.

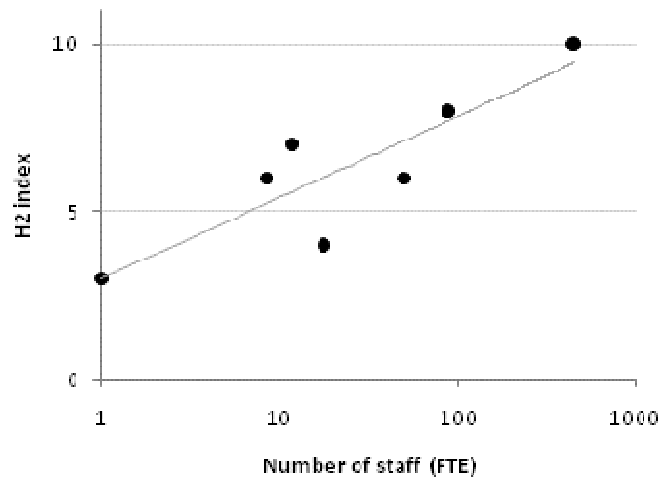

**Figure 6**. The effect of team size of $H_2$.

**Conclusion**

Some of the bibliometric data (especially mean RCI) collated by the ERA and tabulated for presentation to Research Evaluation Committees (RECs) are potentially problematic. One issue is the use of mean RCIs despite the statistical issues associated with skewed distributions. Another limitation is that the computation of percentiles on a year-by-year basis places undue weight on most recent publications, creating an opportunity for manipulation. The tabular data comprising RCIs and percentiles offer the RECs great detail but little synthesis, and it may be more effective to offer additional syntheses such as the indirect $H_2$ which may be helpful to RECs. The indirect $H_2$ index is objective, can be computed automatically and efficiently from databases such as Scopus, is resistant to manipulation, and a good predictor of impact (in the sense pursued through ERA). However, efficient calculation of indices such as $H_2$ requires revision of both the number and diversity of the Fields of Research (FoRs), and of the multiple FoRs assigned to journals within the ERA system.

Since $H_2$ is a newly introduced indicator it is important that further studies on its usefulness and validity should be conducted. Independent of the results of these studies, it is always necessary that this indicator is used in combination with standard bibliometric measures (e.g., RCI classes). Research performance is a multi-dimensional issue which should be reflected by more than one indicator.

# References

Abramo, G., D'Angelo, C. A., & Caprasecca, A. (2009). Allocative efficiency in public research funding: Can bibliometrics help? *Research Policy* 38(1), 206–215.

Abramo, G. & D'Angelo, C.A. (2011). Evaluating research: from informed peer review to Bibliometrics. *Scientometrics* 87, 499-514.

Alonso, S., Cabrerizo, F.J., Herrera-Viedma, E. & Herrera F. (2009). h-Index: A review focused in its variants, computation and standardization for different scientific fields. *Journal of Informetrics* 3, 273-289.

Anon. (2011). List of forestry universities and colleges. Wikipedia, http://en.wikipedia.org/wiki/List_of_forestry_universities_and_colleges [5 August 2011]

ANZSRC (2008). Australian and New Zealand Standard Research Classification. http://www.abs.gov.au/Ausstats/abs@.nsf/0/6BB427AB9696C225CA2574180004463E [5 August 2011].

ARC (2010a). The Excellence in Research for Australia (ERA) Initiative. http://www.arc.gov.au/era/default.htm [5 August 2011].

ARC (2010b). Ranked Journal List. Australian Research Council, http://www.arc.gov.au/xls/ERA2010_journal_title_list.xls [5 August 2011].

ARC (2011a). What is the change to the journal indicator? Australian Research Council. http://www.arc.gov.au/era/faq.htm#2012 [5 August 2011].

ARC (2011b). ERA 2010 National Report. Australian Research Council, 320 pp. http://www.arc.gov.au/pdf/ERA_report.pdf [5 August 2011]

ARC (2011c). ERA 2010 Citation Benchmark Methodology. Australian Research Council, 21 pp, http://www.arc.gov.au/pdf/era_2010_citation_benchmark_methods.pdf [5 August 2011]

ARC (2011e). ERA 2010 results by Field of Research Code. Australian Research Council. http://www.arc.gov.au/era/outcomes_2010/FoRindex [5 August 2011]

Arencibia-Jorge, R., Barrios-Almaguer, I., Fernández-Hernández, S., Carvajal-Espino, R. (2008). Successive H indices and its applying in the institutional evaluation: a case study. J*ournal of the American Society for Information Science and Technology* 59(1), 155–157.

Atkinson, R. & McLoughlin, C. (2011). Dawn of a new ERA? *Australasian Journal of Educational Technology* 27(3), iii-viii. http://ascilite.org.au/ajet/ajet27/editorial27-3.html

Bornmann, L. (2010). Mimicry in science? *Scientometrics*, 86(1), 173-177.

Bornmann, L. (2011). Scientific peer review. *Annual Review of Information Science and Technology*, 45, 199-245.

Bornmann, L. & Mutz, R. (2008). Are there better indices for evaluation purposes than the h index? A comparison of nine different variants of the h index using data from Biomedicine. *Journal of the American Society for Information Science and Technology* 59(5), 830–837.

Bornmann, L., de Moya Anegón, F., & Leydesdorff, L. (2010). Do scientific advancements lean on the shoulders of giants? A bibliometric investigation of the Ortega hypothesis. *PLoS ONE*, 5(10), e13327, doi:10.1371/journal.pone.0013327.

Bornmann, L., Mutz, R., Hug, S.E., Daniel, H.-D. (2011). A multilevel meta-analysis of studies reporting correlations between the h index and 37 different h index variants. *Journal of Informetrics* 5, 346–359.

Bornmann, L., Schier, H., Marx, W., & Daniel, H.-D. (in press). Does the h index for assessing single publications really work? A case study on papers published in chemistry. *Scientometrics*.

Carr, K. (2009). A message from the Minister. Australian Research Council, Discovery Newsletter, Autumn 2009, http://www.arc.gov.au/rtf/discovery_autumn09.rtf

Cooper, S. & Poletti, A. (2011). The new ERA of journal ranking. *Australian Universities' Review* 53, 57-65.

Donovan, C. & Butler, L. (2007). Testing novel quantitative indicators of research 'quality', esteem and 'user engagement': an economics pilot study. *Research Evaluation* 16(4), 231–242.

Egghe, L. (2008). Modelling successive h-indices. *Scientometrics* 77(3), 377–387.

Egghe, L. (2011). The single publication H-index and the indirect H-index of a researcher. *Scientometrics* 88(3), 1003-1004.

Falagas, M.E., Pitsouni, E.I., Malietzis, G.A. & Pappas, G. (2008). Comparison of PubMed, Scopus, Web of Science, and Google Scholar: strengths and weaknesses. *The FASEB Journal* 22, 338-342.

Haddow, G. & Genoni, P. (2010). Citation analysis and peer ranking of Australian social science journals. *Scientometrics* 85, 471-487.

Hicks, D. (2009). Evolving regimes of multi-university research evaluation. *High Educ* 57, 393–404.

Hirsch, J.E. (2005) An index to quantify an individual's scientific research output. *Proceedings of the National Academy of Sciences* 102, 16569–16572.

Jones, D. (2011). Excellence in Research for Australia – a new ERA. *Research Trends* 23, 10-11. http://www.researchtrends.com/wp-content/uploads/2011/06/Research_Trends_Issue23.pdf

Labbé, C. (2010). Ike Antkare, one of the great stars in the scientific firmament. *ISSI Newsletter* 6(2), 48-52.

Lamp, J. (2011). 2010 finalised journals in an ANZ Field of Research: 0705 Forestry Sciences. http://lamp.infosys.deakin.edu.au/era/?page=jfordet10&selfor=0705 [5 August 2011]

Laband, D.N. & Zhang, D. (2006). Citations, publications, and perceptions-based rankings of the research impact of North American forestry programs. *Journal of Forestry* 104(5), 254-162.

Leydesdorff, L., Bornmann, L., Mutz, R., & Opthof, T. (2011). Turning the tables in citation analysis one more time: principles for comparing sets of documents. *Journal of the American Society of Information Science and Technology*, 62(7), 1370-1381.

Li, J., Burnham, J.F., Lemley, T. & Britton, R.M. (2010). Citation Analysis: Comparison of Web of Science, Scopus, SciFinder, and Google Scholar. *Journal of Electronic Resources in Medical Libraries* 7(3), 196-217.

Meho, L.I. & Yang, K. (2007). Impact of data sources on citation counts and rankings of LIS faculty: Web of science versus scopus and google scholar. *Journal of the American Society for Information Science and Technology* 58, 2105–2125.

Moed, H.F. (2009). New developments in the use of citation analysis in research evaluation. *Arch. Immunol. Ther. Exp.* 57, 13–18.

Moosa, I. (2011). The demise of the ARC journal ranking scheme: an ex post analysis of the accounting and finance journals. Accounting & Finance 51(3), 809-836.

Moya-Anegón, F.d., Chinchilla-Rodríguez, Z., Vargas-Quesada, B., Corera-Álvarez, E., Muñoz-Fernández, F.J., González-Molina, A. & Herrero-Solana, V. (2007). Coverage analysis of Scopus: A journal metric approach. *Scientometrics* 73, 53–78.

Norris, M. & Oppenheim, C. (2010). "The h-index: a broad review of a new bibliometric indicator", Journal of Documentation, 66(5), 681 – 705.

Northcott, D. & Linacre, S. (2010). Producing Spaces for Academic Discourse: The Impact of Research Assessment Exercises and Journal Quality Rankings. *Australian Accounting Review* 20, 38-54.

Pontille, D., & Torny, D. (2010). The controversial policies of journal ratings: Evaluating social sciences and humanities. *Research Evaluation* 15, 347-360.

Radicchi, F., & Castellano, C. (2011). Rescaling citations of publications in physics. *Physical Review E*, 83(4). doi: 10.1103/PhysRevE.83.046116.

Rousseau, R., Yang, L. & Yue, T. (2010). A discussion of Prathap's h2-index for institutional evaluation with an application in the field of HIV infection and therapy. *Journal of Informetrics* 4(2), 175-184.

Ruane, F. & Tol, R.S.J. (2008). Rational (successive) h-indices: An application to economics in the Republic of Ireland. *Scientometrics* 75(2), 395–405.

Runeson, G. (2011). The demise of the journal ranking: a victory for common sense. *Australasian Journal of Construction Economics and Building* 11 (2), 99-100.

Schreiber, M. (2010). Twenty Hirsch index variants and other indicators giving more or less preference to highly cited papers. *Annalen der Physik* 522(8), 536–554.

Schubert, A. (2007) Successive i-indices. Scientometrics 70(1), 201-205.

Schubert, A. (2009). Using the h-index for assessing single publications. *Scientometrics* 78(3), 559-565.

Serenko, A. & Dohan, M. (2011). Comparing the expert survey and citation impact journal ranking methods: Example from the field of Artificial Intelligence. *Journal of Informetrics* 5(4), 629-648.

Sinclair, R. (2010). RD&E strategy for the forest and wood products sector. Forest & Wood Products Australia, 57 pp. www.daff.gov.au/__data/assets/word_doc/0010/1770751/forest.doc

Singh, G., Haddad, K.M., & Chow, C.W. (2007). Are articles in "top" management journals necessarily of higher quality? *Journal of Management Inquiry*, 16(4), 319–331.

Slyder, J.B., Stein, B.R., Sams, B.S., Walker, D.M., Beale, B.J., Feldhaus, J.J., Copenheaver, C.A. (2011). Citation pattern and lifespan: a comparison of discipline, institution and individual. *Scientometrics* 89(3), 955-966.

Svantesson, D.J.B. & White, P. (2009). Entering an era of research ranking - will innovation and diversity survive? *Bond Law Review* 21(3), 7.

Taylor, J. (2011). The Assessment of Research Quality in UK Universities: Peer Review or Metrics? *British Journal of Management* 22, 202–217.

Thor, A. & Bornmann, L. (2011). The calculation of the single publication h index and related performance measures: A web application based on Google Scholar data. Online Information Review 35(2), 291-300. (Utility available here: http://labs.dbs.uni-leipzig.de/gsh/ ).

Vanclay, J.K. (2008). Ranking forestry journals using the h-index. *Journal of Informetrics* 2, 326-334.

Vanclay, J.K. (2011). An evaluation of the Australian Research Council's journal ranking. *Journal of Informetrics* 5, 265-274.

Vanclay, J.K. (2012). What was wrong with Australia's journal ranking? *Journal of Informetrics* 6, 53– 54.

**Annex 1. List of journals included within 0705 Forestry Sciences in ERA 2010 (ARC 2010b, Lamp 2011)**

Acta Facultatis Xylologiae
Acta Scientiarum Polonorum
Agricultural and Forest Meteorology
Agroforestry Systems
Annales de la Recherche Forestiere Au Maroc
Annals of Forest Science
Annals of Forestry
Appita Journal
Arboricultural Journal
Australian Forestry
Austrian Journal of Forest Science
Baltic Forestry
Bangladesh Journal of Forest Science
Bois et Forets des Tropiques
Canadian Journal of Forest Research
Centralblatt fuer Das Gesamte Forstwesen
Cerne
Ciencia Florestal
Croatian Journal of Forest Engineering
Drewno
European Journal of Forest Research
European Journal of Wood and Wood Industries
Floresta
Folia Forestalia Polonica. Series A. Lesnictwo
Folia Oecologica
Forest Biometry Modelling and Information Sciences
Forest Ecology and Management
Forest Pathology
Forest Products Journal
Forest Science
Forest Usufructs
Forestry Chronicle
Forestry Studies in China
Forestry
Forests, Trees and Livelihoods
Frontiers of Forestry in China
Glasnik za Sumske Pokuse
Holzforschung
IAWA Journal
Indian Forester
Indian Journal of Forestry
International Forestry Review
International Journal of Forest Engineering
International Journal of Forest Genetics
International Journal of Wildland Fire
Investigacion Agraria. Sistemas y Recursos Forestales
Istanbul Universitesi Orman Fakultesi Dergisi Seri A
Jiangsu Linye Keji
Journal of Forest Research
Journal of Forest Science-Prague
Journal of Forestry
Journal of Pulp and Paper Science
Journal of Sustainable Forestry
Journal of Tropical Forest Science
Journal of Wood Science
Linye Kexue
Metstieteen Aikakauskirja
Nederlands Bosbouw Tijdschrift
New Forests
New Zealand Journal of Forestry Science
Paper360
Revista Forestal Latinoamericana
Revista Padurilor: silvicultura si exploatarea padurilor
Scandinavian Journal of Forest Research
Schweizerische Zeitschrift fuer Forstwesen
Scientia Forestalis
Scottish Forestry
Silva Fennica
Silva Lusitana
Silvae Genetica
Small-Scale Forestry
Southern Forests
Southern Journal of Applied Forestry
Sumarski List
Sylwan: czasopismo lesne
Tasforests
Tree and Forestry Science and Biotechnology
Tree Genetics and Genomes
Tree Physiology
Tree-Ring Research
Trees: structure and function
Urban Forestry and Urban Greening
Western Journal of Applied Forestry
Wood Science and Technology
Zhejiang Linxueyuan Xuebao

Annex 2

| H₂ | Higher Educational Institution |
|---|---|
| 9+ | **USA**: UC Berkeley, Oregon State; **Canada**: UBC |
| 8 | **Finland**: U Helsinki, U Eastern Finland; **Italy**: Viterbo; **Sweden**: SLU; **USA**: Harvard, New Hampshire |
| 7 | **Australia**: ANU; **Belgium**: Antwerp; **Canada**: Laval, McGill, Quebec, Toronto; **Estonia**: U Tartu; **Germany**: Technische U Munchen; **Netherlands**: Wageningen; **USA**: Alaska, Duke, Minnesota, Montana, NAU, Idaho, Washington, Wisconsin Madison, Yale |
| 6 | **Australia**: JCU, Melbourne, SCU, UWA; **Austria**: BOKU; **Belgium**: KU Leuven; **Canada**: Alberta, New Brunswick; **Germany**: Freiburg, Göttingen; **Italy**: Padova; **Japan**: Gifu, Tsukuba; **Portugal**: ISA Lisbon; **Switzerland**: ETH; **USA**: Colorado, Cornell, Florida, Georgia, Maine, Michigan State, Michigan Tech, NC State, Ohio State, UC Davis, Maryland, Vermont, Virginia Tech. |
| 5 | **Australia**: UQ; **Belgium**: Gembloux, Gent; **Brazil**: Brasilia; **Canada**: Winnipeg; **China**: NE Forestry U; **Czech**: Mendel; **France**: Henri Poincare, Montpellier; **Germany**: Bayreuth, Hamburg; **Japan**: Hokkaido, Kyoto, Kyushu Nagoya, Tokyo U A&T, Tokyo; **Mexico**: Santiago de Compostella; **Norway**: U Miljø- og Biovitenskap; **Slovenia**: Ljubljana; **Sweden**: Lund; **UK**: Edinburgh, Leeds, Oxford; **USA**: Clemson, Penn State, Purdue, SUNY, Texas A&M, Illinois, Massachusetts, Missouri, Washington State. |
| 4 | **Argentina**: Beunos Aries; **Australia**: Monash U, Murdoch U, U Tasmania; **Belgium**: Vrije U Brussel; **Brazil**: U Sao Paulo; **Canada**: Lakehead U, UNBC, U Victoria; **Chile**: Conception, U Austral; **China**: Graduate U Chinese Academy of Sciences; **Denmark**: Copenhagen U; **Finland**: Abo Akademi, Jyväskylän, Oulun; **France**: U Bordeaux, U Paris-Sud; **Germany**: TU Dresden; **Hungary**: Eötvös Loránd U; **Ireland**: U College Cork; **Italy**: Firenze, U Napoli Federico II; **Netherlands**: Utrecht U; **Norway**: Norges Teknisk-Naturvitenskapelige U; **NZ**: U Canterbury; **Portugal**: U Tras-o-Montes, U Aveiro; **South Africa**: Pretoria; **Spain**: U Leon, U Cordoba, U Extremadura, U Valladolid, U Politecnica Madrid, U Lleida, U Politecnica Valencia; **Sweden**: Umea U; **Thailand**: Kasetsart; **UK**: Aberdeen, Bangor, Reading; **USA**: Auburn, Iowa State, Louisiana State, Mississippi State, Oklahoma State, Southern Illinois U Carbondale, U Arkansas Monticello, U Kentucky, U Tennessee, U Guelph, Utah State U, West Virginia U |
| 3 | **Argentina**: C R U Baroloche, U N La Plata, U N Patagonia; **Australia**: Curtin U, Griffith U; **Austria**: TU Innsbruck; **Brazil**: U Federal Santa Maria, U Federal de Vicosa , U Federal do Parana; **Canada**: Moncton; **Chile**: Pontificia U Católica de Chile, U Chile; **China**: Bejing Forestry U, China Agric U, Fujian Agric & Forestry U, Huazhong Agric U, Northwest A&F U; **Costa Rica**: CATIE; **Croatia**: U Zagreb; **Czech**: Czech U Life Sciences Prague; **Estonia**: Estonian U Life Sciences; **Ethiopia**: Wondo Genet College of Forestry; **Finland**: Aalto U; **France**: Universite Blaise Pascal; **Greece**: Technologiko Ekpaideftiko Idryma Kavala; **Ireland**: U College Dublin; **Italy**: U Molise, U Basilicata, U Palermo; **Israel**: Hebrew U Jerusalem; **Japan**: Ehime U, Shimane U; **Malaysia**: U Malaysia Sabah, U Putra Malaysia; **Mexico**: U Michoacana de San Nicolás de Hidalgo, U Nacional Autónoma de México; **Norway**: U Oslo; **NZ**: Lincoln U; **Poland**: U Przyrodniczy w Poznaniu; **Portugal**: U Coimbra; **Russia**: Sukachev Institute of Forest; **Singapore**: NUS; **South Africa**: U Stellenbosch, U KwaZulu-Natal; **South Korea**: Kangwon National U, Seoul National U; **Sweden**: Luleå tekniska U, Växjö universitet; **Taiwan**: N Taiwan U, National Chung Hsing University; **Turkey**: U Istanbul; **UK**: Cambridge, Napier U; **USA**: Humboldt State U; **Venezuela**: U Los Andes |
| 2 | **Benin**: U Abomey-Calavi; **Brazil**: UNESP-U Estadual Paulista; **Bosnia**: U Sarajevo; **Chile**: U de la Frontera; **China**: Central South U Forestry & Technology, Hebei Agric U, Nanjing Forestry U, South China Agric U, Zhejiang Forestry U; **Congo**: Universite Marien Ngouabi; **Costa Rica**: Instituto Tecnológico de Costa Rica; **Cuba**: U Pinar del Rio ; **Denmark**: Aarhus Universitet; **Ecuador**: U Nacional de Loja; **Ethiopia**: U Gondar; **France**: Nancy U; **Germany**: Ludwig-Maximilians-U München, U Appl Sciences Rottenburg; **Ghana**: Kwame Nkrumah U; **Greece**: Aristotle U Thessaloniki, Dimokrition Panepistimion Thrakis; **Honduras**: Escuela Nacional de Ciencias Forestales; **Hungary**: U West Hungary; **Iceland**: Agric U Iceland; **India**: Banaras Hindu U, Kerala Agric U; **Indonesia**: Bogor Agric U, Gadja Mada, U Padjadjaran, U Palangka Raya; **Iran**: Daneshgahe Tarbiat Modares, U Tehran; **Italy**: U Torino, U Reggio Calabria; **Japan**: Iwate U, Shinshu U, Tottori U, U Kobe, U Miyazaki, Utsunomiya U; **Kenya**: Moi U; **Morocco**: Faculté des Sciences Semlalia; **Mexico**: U Autónoma de Chapingo, U Autónoma de Nuevo León, U Juárez del Estado de Durango; **Nicaragua**: U Nacional Agraria; **NZ**: U Waikato; **Poland**: Warsaw U Life Sciences, U Agric Krakow; **Romania**: U Transilvania, U Transilvania din Brasov; **Russia**: Saint Petersburg State Forest Technical Academy; **Serbia**: U Belgrade; **Slovakia**: TU Zvolen; **South Africa**: U Witwatersrand; **Sri Lanka:** U Peridenia; **Tanzania**: Sokoine: U; **Thailand:** Mahidol U; **Turkey**: Abant Izzet Baysal U, Kafkas U, Kahramanmaras Sütçü Imam U, Karadeniz Teknik U, Kastamonu U, Süleyman Demirel U; **Uganda**: Makerere U; **UK**: U Nottingham; **USA**: Louisiana Tech, Stephen F. Austin State U; **Vietnam**: Hue U Agric and Forestry; |
| 1 | **Argentina**: U Nacional de Cuyo, U Nacional de Misiones; **Bangladesh**: Khulna U, Shahjalal U Science and Technology, U Chittagong; **Bolivia**: U Autónoma Gabriel René Moreno, U Mayor de San Andres; **Botswana**: U Botswana **Brazil**: U Federal de Lavras; **Bulgaria**: U Forestry; **Burkina Faso**: U Ouagadougou; **Cameroon**: U Yaoundé; **Chile**: U Talca; **China**: Guizhou U, Sichuan Agric U, Southwest Forestry U; **Colombia**: U Nacional de Colombia Medellin; **Core d'Ivoire**: U Cocody; **Costa Rica**: U Nacional; **Czech**: U Karlova v Praze; **Finland**: Mikkelin Ammattikorkeakoulu, Pohjois-Karjalan Ammattikorkeakoulu; **France**: Ecole Nationale Supérieure Agronomique de Toulouse, U Claude Bernard Lyon; **Germany**: U Applied Sciences Rosenheim; **Guatemala**: U San Carlos; **Guiana**: Université Antilles-Guyane; **India**: Aligarh Muslim U, Dr Yashwant Singh Parmar U, Hemwati Nandan Bahuguna Garhwal U, Jai Narain Vyas U, North-Eastern Hill U, Punjab Agric U, Rajendra Agric U; **Indonesia**: Tanjungpura U, Winaya Mukti U; **Iran**: Gorgan U Agric Sciences and Natural Resources, U Guilan, U Kurdistan; **Jordan**: U Science and Technology; **Laos**: National U of Laos; **Latvia**: U Latvia; **Lebanon**: American U Beirut; **Lithuania**: Lithuanian U Agric; **Macedonia**: SS Cyril and Methodius U; **Malawi**: Mzuzu U, U Malawi; **Malaysia**: U Malaysia Sarawak, U Sains Malaysia, U Teknologi MARA; **Mexico**: U Guadalajara; **Morocco**: U Cadi Ayyad; **Mozambique**: U Eduardo Mondlane; **Nepal**: Tribhuvan U; **Nigeria**: Federal U Technology, U Ibadan, U Ado-Ekiti, U Agric Abeokuta; **Pakistan**: U Peshawar; **Peru**: U Nacional Agraria La Molina, U Nacional de la Amazonia Peruana Iquitos; **Philippines**: U Philippines Los Banos, Visayas State U; **Poland**: U Warszawski; **Russia**: Moscow State Forest U; **Senegal**: U Cheikh Anta Diop; **South Africa**: Nelson Mandela Metropolitan U; **Sri Lanka**: U Ruhuna; **Tanzania**: U Dar Es Salaam; **Thailand**: Chiang Mai U; **Trinidad and Tobago**: U West Indies; **Tunisia**: El Manar U; **Turkey**: Artvin Coruh U, Black Sea Technical U, Zonguldak Karaelmas U; **UK**: Central Lancashire, Wolverhampton; **Ukraine**: Ivan Franko National U L'viv; **Uruguay**: U de la Republica; **USA**: Alabama A&M U, Cal Poly, U Wisconsin Stevens Pt; **Zambia**: Copperbelt U, U Zambia; **Zimbabwe**: U Zimbabwe. |
| 0 | **Albania**: Agric U Tirana; **Algeria**: U Tlemcen; **Bolivia**: U Autónoma de Beni, U Autónoma Juan Misael Saracho; **Brazil**: U Federal do Espirito Santo; **Burkina Faso**: U Polytechnique de Bobo-Dioulasso; **Cameroon**: U Ngaoundéré; **China**: Henan Agricultural U; **Ecuador**: U Técnica Estatal de Quevedo; **Egypt**: Alexandria U; **Finland**: Seinäjoki U Applied Sciences; **Honduras**: Escuela Nacional de Ciencias Forestales; **Indonesia**: Bengkulu U, Mulawarman U, U Lampung; **Iran**: Sari Agricultural and Natural Resources U, Urmia U; **Nigeria**: Delta State U, U Uyo; **Peru**: U Agraria de la Selva, U Federal Rural da Amazonia; **Philippines**: Isabela State U, Mariano Marcos State U; **PNG**: U Papua New Guinea; **South Korea**: Yeungnam U; **Spain**: U Católica de Ávila; **Sudan**: U Khartoum; **Ukraine**: Ukrainian State Forestry Engineering U; **Vietnam**: U Natural Sciences. |